\documentclass{PoS}

\newcommand{\nn}{\nonumber}
\newcommand{\ba}{\begin{eqnarray}}
\newcommand{\ea}{\end{eqnarray}}
\newcommand{\bd}{\begin{displaymath}}
\newcommand{\ed}{\end{displaymath}}
\newcommand{\be}{\begin{equation}}
\newcommand{\ee}{\end{equation}}

\title{A new moving frame to extract scattering phases in lattice QCD}

\ShortTitle{A new moving frame to extract scattering phases in lattice QCD}

\author{
\vspace{-0.05\textheight}
\begin{flushleft}
\hspace{0.75\textwidth}DESY 10-177\\
\hspace{0.75\textwidth}MS-TP-10-15\\
\hspace{0.75\textwidth}KEK-CP-244\\
\hspace{0.75\textwidth}JLAB-THY-11-1349
\end{flushleft}
        \speaker{Xu Feng}\thanks{Current address:\ KEK.}\,\,$^{a,b}$,
        Karl Jansen$^a$ and Dru B. Renner\thanks{Current address:\ Jefferson Lab.}\,\,$^a$\\
        \llap{$^a$}NIC, DESY, Platanenallee 6, D-15738 Zeuthen, Germany\\
        \llap{$^b$}Universit\"at M\"unster, Institut f\"ur Theoretische Physik, Wilhelm-Klemm-Strasse 9, D-48149 M\"unster, Germany\\
        E-mail: \email{xufeng@post.kek.jp}}

\abstract{We present a derivation of the finite-size formulae in a moving frame with 
total momentum $\mathbf P=(2\pi/L)({\mathbf e}_1+{\mathbf e}_2)$. These formulae allow 
us to calculate the S-wave and P-wave scattering phases 
at more energies with a fixed lattice size and thus 
help us to determine the resonance parameters precisely.
\vspace{40pt}
\begin{center}
\includegraphics[width=100pt]{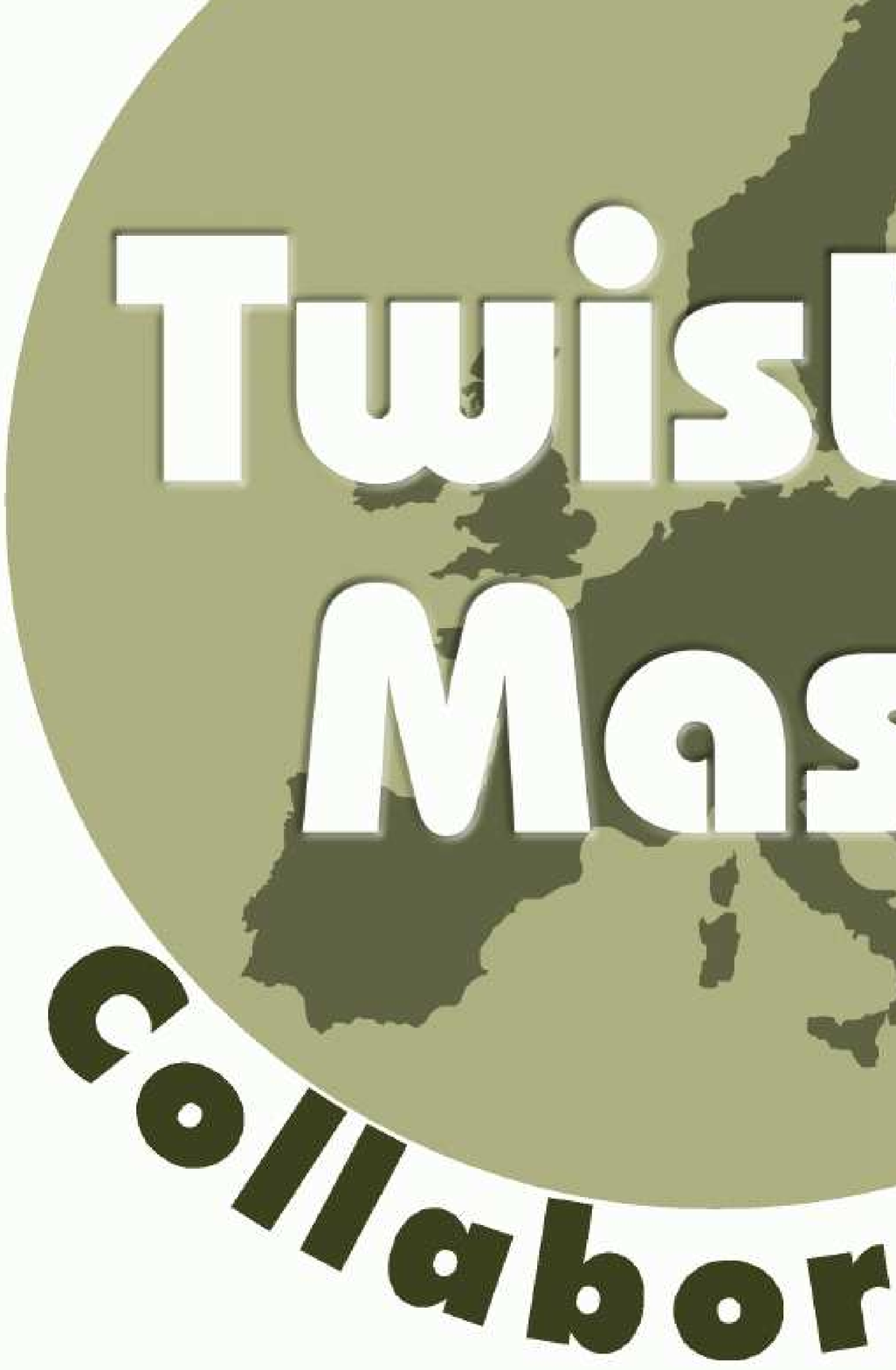}
\end{center}
}

\FullConference{The XXVIII International Symposium on Lattice Field Theory, Lattice2010\\
		June 14-19, 2010\\
		Villasimius, Italy}

\begin{document}
\section{Introduction}

The finite-size method, originally proposed by L\"uscher~\cite{Luscher:1990ux}, allows us to calculate the 
scattering phase of two particles in infinite volume
from the energy eigenvalues of a two-particle system enclosed in
a finite spatial box.
For a fixed finite volume, the scattering phase is determined at a discrete set of energies.
To better capture the energy dependence of the scattering phase, we would like to increase the number of
energy eigenvalues accessible to lattice calculations. Particularly, in a scattering channel where a resonance
appears, we want to calculate the scattering phase at many energies in 
the resonance region in order to determine the resonance parameters such as the mass and width.

One method to address this issue is to use different lattice sizes to obtain the scattering phase at more
energy values.
However, this requires additional simulations that are increasingly more demanding at large volumes.
An alternative way is to use the moving frame (MF) technique, introduced by Rummukainen and Gottlieb~\cite{Rummukainen:1995vs},
which generalizes L{\"u}scher's original method from the center-of-mass frame (CMF) to a MF.
The key point is that the energy spectrum calculated in a MF is different from the
one obtained in the CMF. Thus combining the CMF and a MF allows us to compute the
scattering phase at more energies using the same lattice size.  This will increase the accuracy of the
calculation of the desired resonance parameters for less computational cost.

As an application, we have performed a calculation of the P-wave pion-pion scattering phase in
the $\rho$-meson decay channel using three Lorentz frames: the CMF, the original MF with 
total momentum ${\mathbf P}=(2\pi/L)\,{\mathbf e}_3$ (MF1)
and a second MF with ${\mathbf P}=(2\pi/L)({\mathbf e}_1+{\mathbf e}_2)$ (MF2).
This approach allowed us to determine the scattering phase as a function of the
energy covering the resonance region~\cite{Feng:2010es}.
MF2 is a new MF for which the finite-size formulae were not yet available
in the literature.
In this work,  we present a derivation of the finite-size formulae for this MF. Our derivation closely follows
the work in Refs.~\cite{Luscher:1990ux,Rummukainen:1995vs}.    

\section{Two-particle wave function}
 Consider a two-particle system with total momentum ${\mathbf P}\neq\mathbf 0$. 
 The state of such a system can be described by 
 a wave function $\phi(\mathbf x)$, where $\mathbf x$ is the relative position between the two particles.
 To establish the formula for the scattering phase, which is naturally defined in the CMF, we need to transform 
 the scattering system from the MF to the CMF. 
 This is accomplished by the appropriate Lorentz transformation, parameterized by $\vec{\gamma}$, under which the wave function transforms from 
 $\phi$ to $\phi_{CM}$.
 
 We assume that the two-particle interaction vanishes 
 in the region where $|{\mathbf x}|>R$, called the exterior region. Then in the exterior region  
 $\phi_{CM}({\mathbf x})$ satisfies the Helmholtz equation
 \be
 \label{eq:helmholtz}
 (\nabla^2+p^2)\phi_{CM}({\mathbf x})=0\;,\quad\mathrm{for\;\;}|{\mathbf x}|>R\;,
 \ee
 where $p$ is given by the energy-momentum relation $p^2=(E_{CM}/2)^2-m^2$ with 
 $E_{CM}$ the center-of-mass energy and $m$ the single-particle mass.
 If we consider such a two-particle system enclosed in a box with finite size $L>2R$, then 
 the total momentum ${\mathbf P}$ 
 is discretized as ${\mathbf P}=(2\pi/L)\,{\mathbf d}$ with ${\mathbf d}\in\mathbb{Z}^3$ and
 $\phi_{CM}({\mathbf x})$ satisfies the ${\mathbf d}$-periodic
 boundary condition
 \be
 \label{eq:d_boundary_cond}
 \phi_{CM}({\mathbf x})=(-1)^{{\mathbf d}\cdot{\mathbf n}}\phi_{CM}({\mathbf x}+\vec{\gamma}{\mathbf n}L)\;,
 \quad \mathrm{for\;\;all\;\;}
 {\mathbf n}\in\mathbb{Z}^3\;.
 \ee 
 The details of the derivations of Eqs.~(\ref{eq:helmholtz}) and (\ref{eq:d_boundary_cond}), along with any unexplained
 notation such as $\vec{\gamma}{\mathbf n}$, can be 
 found in Ref.~\cite{Rummukainen:1995vs}.

 A function that satisfies the Helmholtz equation~(\ref{eq:helmholtz}) and obeys the
 ${\mathbf d}$-periodic boundary condition~(\ref{eq:d_boundary_cond}) is called a
  singular ${\mathbf d}$-periodic solution of the Helmholtz equation.
 A simple example of such a solution is the Green's function
 \bd
 G^{{\mathbf d}}({\mathbf x},p^2)=\gamma^{-1}L^{-3}\sum_{{\mathbf k}\in P_{{\mathbf d}}}
 \frac{e^{i{\mathbf k}\cdot{\mathbf x}}}{k^2-p^2}\;,
 \ed
 where the summation of the momentum runs over
 \bd
 P_{{\mathbf d}}=\left\{{\mathbf k}\in\mathbb{R}^3\left|\;{\mathbf k}=\frac{2\pi}{L}
 \vec{\gamma}^{-1}({\mathbf m}+\frac{1}{2}{\mathbf d})\;,\quad \textmd{for}\;\;
 {\mathbf m}\in\mathbb{Z}^3\right.\right\}\;.
 \ed
 More singular ${\mathbf d}$-periodic solutions can be generated from the Green's function
 by introducing the harmonic polynomials
 \be
 \label{eq:harmonic_poly}
 \mathcal{Y}_{lm}({\mathbf x})=r^lY_{lm}(\hat{\mathbf x})\;,\quad r=|\mathbf x|\;,
 \ee
 and defining
 \bd
 G_{lm}^{\mathbf d}({\mathbf x},p^2)=\mathcal{Y}_{lm}({\mathbf \nabla})G^{{\mathbf d}}({\mathbf x},p^2)\;.
 \ed
 In Eq.~(\ref{eq:harmonic_poly}) the notation $\hat{\mathbf x}$ represents the solid angle parameters $(\theta,\varphi)$ of $\mathbf x$ 
 in spherical coordinates.
 It was proved in Ref.~\cite{Luscher:1990ux} that the $G_{lm}^{\mathbf d}({\mathbf x},p^2)$ are complete and linearly
 independent. Therefore $\phi_{CM}({\mathbf x})$ can be expanded in terms of $G_{lm}^{\mathbf d}({\mathbf x},p^2)$ as
 \bd
 \label{wave_func_finite}
 \phi_{CM}({\mathbf x})=\sum_{l,m}\nu_{lm}G_{lm}^{\mathbf d}({\mathbf x},p^2)\;,\quad\mathrm{for\;\;}R<|{\mathbf x}|<L/2\;.
 \ed

 As given in Ref.~\cite{Rummukainen:1995vs}, $G_{lm}^{\mathbf d}({\mathbf x},p^2)$ can be expanded in terms of 
 the spherical harmonics $Y_{lm}(\hat{\mathbf x})$ 
 and spherical Bessel functions $j_l(pr)$ and $n_l(pr)$ through
 \be
 \label{eq:expansion_Bessel_L}
 G_{lm}^{\mathbf d}({\mathbf x},p^2)=\frac{(-1)^l}{4\pi}p^{l+1}\left\{Y_{lm}(\hat{\mathbf x})n_l(pr)+\sum_{l',m'}
 \mathcal{M}^{\mathbf d}_{lm,l'm'}(p)Y_{l'm'}(\hat{\mathbf x})j_{l'}(pr)\right\}\;.
 \ee
 Due to symmetry considerations, some $\mathcal{M}^{\mathbf d}_{lm,l'm'}(p)$ vanish. 
 For the MF2 (${\mathbf d}={\mathbf e}_1+{\mathbf e}_2$), we list all the non-zero values of $\mathcal{M}^{\mathbf d}_{lm,l'm'}(p)$ for $l,l'=0,1$ in Table~\ref{value_M}. There we expand $\mathcal{M}^{\mathbf d}_{lm,l'm'}(p)$
 in terms of the modified zeta function, which is defined by
 \bd
 \mathcal{Z}_{lm}^{\mathbf d}(s;q^2)=\sum_{\frac{2\pi}{L}{\mathbf n}\in P_{\mathbf d}}\frac{\mathcal{Y}_{lm}^*({\mathbf n})}{\left(|{\mathbf n}|^2-q^2\right)^{s}}\;,\quad q=\frac{pL}{2\pi}\;.
 \ed
 This zeta function is formally divergent and needs to be analytically continued. Ref.~\cite{Rummukainen:1995vs} 
 gives a numerically calculable expression of $\mathcal{Z}_{lm}^{\mathbf d}(1;q^2)$, which is not, however, valid
 for the MF2. Here we give a more general expression
 that can be used in all the different MFs
 \ba
\mathcal{Z}_{lm}^{{\mathbf d}}(1;q^2)&=&\gamma\int_0^{1}dte^{tq^2}\sum_{{\mathbf u}\in\mathbb{Z}^3,{\mathbf u}\neq{\mathbf 0}}
(-1)^{{\mathbf u}\cdot{\mathbf d}}
i^l
\mathcal{Y}_{lm}^*(-\frac{\pi\vec{\gamma}{\mathbf u}}{t})(\frac{\pi}{t})^{3/2}
\exp(-\frac{|\pi\vec{\gamma}{\mathbf u}|^2}{t})\nn\\
&+&\gamma\int_0^{1}dt(e^{tq^2}-1)
\frac{1}{\sqrt{4\pi}}\delta_{l0}\delta_{m0}(\frac{\pi}{t})^{3/2}
-\gamma\pi\delta_{l0}\delta_{m0}\nn\\
&+&\sum_{\frac{2\pi}{L}{\mathbf n}\in P_{{\mathbf d}}}
\frac{\mathcal{Y}_{lm}^*({\mathbf n})}{|{\mathbf n}|^2-q^2}\exp(-(|{\mathbf n}|^2-q^2))\;.\nn
\ea

\begin{table}[htb]
\begin{center}
\begin{tabular}{|c|c|c|}
  \hline
  $lm$ & $l'm'$ & $\gamma \pi^{3/2}q\mathcal{M}^{{\mathbf d}}_{lm,l'm'}(p)$ \\
  \hline
  00 & 00 & $\mathcal{Z}_{00}^{{\mathbf d}}(1;q^2)$ \\
  \hline
  10 & 10 &$\mathcal{Z}_{00}^{{\mathbf d}}(1;q^2)
  +\frac{2q^{-2}}{\sqrt{5}}\mathcal{Z}_{20}^{{\mathbf d}}(1;q^2)$\\
  \hline
  11 & 11 &$\mathcal{Z}_{00}^{{\mathbf d}}(1;q^2)-\frac{q^{-2}}{\sqrt{5}}\mathcal{Z}_{20}^{{\mathbf d}}(1;q^2)$\\
  $1\bar{1}$ & $1\bar{1}$ &$\mathcal{Z}_{00}^{{\mathbf d}}(1;q^2)-\frac{q^{-2}}{\sqrt{5}}\mathcal{Z}_{20}^{{\mathbf d}}(1;q^2)$\\
  \hline
  11 & $1\bar{1}$ & $-\frac{\sqrt{6}q^{-2}}{\sqrt{5}}\mathcal{Z}_{2\bar{2}}^{{\mathbf d}}(1;q^2)$\\
  $1\bar{1}$ & 11 & $-\frac{\sqrt{6}q^{-2}}{\sqrt{5}}\mathcal{Z}_{22}^{{\mathbf d}}(1;q^2)$\\
  \hline
\end{tabular}
\caption{$\mathcal{M}^{{\mathbf d}}_{lm,l'm'}$ expanded in terms of $\mathcal{Z}_{lm}^{{\mathbf d}}(1;q^2)$ for $l,l'=0,1$.  The notation $\bar{m}=-m$ is used.}
\label{value_M}
\end{center}
\end{table}

\section{Deformed symmetry}
 We introduce the group ${\mathbb G}$ as the set of all lattice rotations and reflections $\hat{R}$ 
 that leave the set of $P_{\mathbf d}$ invariant,
 \be
 \label{eq:constraint2}
 {\mathbb G}=\left\{\hat{R}\left|\;\hat{R}{\mathbf k}\in P_{\mathbf d}\;,\;\;\forall\;{\mathbf k}\in P_{\mathbf d}\right.\right\}.
 \ee
 Let $\mathcal H$ be the space of all ${\mathbf d}$-periodic wave functions $\psi({\mathbf x})$.
 For any $\psi(\mathbf x)\in\mathcal H$ and $\hat{R}\in{\mathbb G}$, 
 the transformed wave function $\hat{R}\psi(\mathbf x)\equiv\psi(\hat{R}^{-1}\mathbf x)$ satisfies
 \be
 \label{eq:symmetric}
 \psi(\hat{R}^{-1}\mathbf x)=
 \sum_{l,m}\nu_{lm}\mathcal{Y}_{lm}(\hat{R}^{-1}{\mathbf \nabla})G^{{\mathbf d}}({\mathbf x},p^2)
 =\sum_{l,m,m'}\nu_{lm}D^{(l)}_{mm'}(\hat{R})\mathcal{Y}_{lm'}({\mathbf \nabla})G^{{\mathbf d}}({\mathbf x},p^2)
 \in{\mathcal H}\;.
 \ee
 This shows that $\mathcal H$ (or equivalently the two-particle system) is closed under the group $\mathbb G$.
 $D^{(l)}_{mm'}(\hat{R})$ used in Eq.~(\ref{eq:symmetric}) is the standard finite-dimensional rotation matrix. It originates from
 \bd
 \hat{R}{\mathcal Y}_{lm}(\mathbf x)={\mathcal Y}_{lm}(\hat{R}^{-1}\mathbf x)=
 \sum_{m'=-l}^{l}D_{mm'}^{(l)}(\hat{R}){\mathcal Y}_{lm'}(\mathbf x)\;.
 \ed
 If we consider a vector space ${\mathcal H}_l$ in which the 
 harmonic polynomials ${\mathcal Y}_{lm}(\mathbf x)$ form an orthonormal basis, then $D^{(l)}_{mm'}(\hat{R})$ is simply an
 irreducible representation (irrep) of the rotational group $O(3)$, which describes the group elements $\hat{R}\in O(3)$ 
 in terms of linear transformations acting on the vector spaces ${\mathcal H}_l$.
 If we restrict $\hat{R}$ to $\hat{R}\in\mathbb G$, then $D^{(l)}_{mm'}(\hat{R})$ is reducible and 
 the vector space ${\mathcal H}_l$ can be further decomposed to sub-spaces that are invariant under $\mathbb G$.
  
 In the case of ${\mathbf d}={\mathbf 0}$ ($\gamma=1$), ${\mathbb G}$ is given by the 
 full cubic group $O_h$. For the MF cases,
 the constraint in Eq.~(\ref{eq:constraint2}) excludes some lattice rotations and 
 ${\mathbb G}$ is reduced to a subgroup of $O_h$. One can prove that Eq.~(\ref{eq:constraint2}) is equivalent to 
 \bd
 {\mathbb G}=\left\{\hat{R}\in O_h\left|\;\hat{R}{\mathbf d}={\mathbf d}\;\;\textmd{or}\;\;\hat{R}{\mathbf d}=-{\mathbf d}
 \right.\right\}\;,
 \ed
 indicating that for ${\mathbf d}\neq{\mathbf 0}$
 ${\mathbb G}$ is the parity doubled little group of ${\mathbf P}=(2\pi/L)\,{\mathbf d}$.  This is the group of rotations that leave the specific direction $\mathbf d$ 
 unchanged combined with reflections of ${\mathbf d}$. 

 In the MF2, $\mathbb G$ is given by the orthorhombic group
 $D_{2h}$, which has 8 one-dimensional irreps: $A^\pm$, $B_1^\pm$, $B_2^\pm$ and $B_3^\pm$.~\footnote{With the change of coordinates $x_3'=\frac{1}{\sqrt{2}}(x_1+x_2)$,
$x_2'=\frac{1}{\sqrt{2}}(x_1-x_2)$ and $x_1'=x_3$, the notation of the irreps used here coincides with the ones used in Ref.~\cite{Landau}, chapter XII.}  
 The index $\pm$ comes from parity, which fixes the transformation behavior of $\phi_{CM}(\mathbf x)$ under 
 reflections ${\mathbf x}\rightarrow-{\mathbf x}$.

 In general, we denote the irreps of $\mathbb G$ as $\Gamma$. As we mentioned,
 the space ${\mathcal H}_l$, which is invariant under $O(3)$, can be decomposed into the irreducible
 sub-spaces ${\mathcal H}_{\Gamma}$ of $D_{2h}$.
 For brevity, we represent these decompositions by only the indices of $l$ and $\Gamma$.
 For $l=0,1,2$ the decompositions are given by
 \ba
 &&l=0\quad\mathrm{decomp}\quad A^+\;,\nn\\
 &&l=1\quad\mathrm{decomp}\quad B_1^- \oplus B_2^- \oplus B_3^-\;,\nn\\
 &&l=2\quad\mathrm{decomp}\quad A^+ \oplus A^+ \oplus B_1^+ \oplus B_2^+ \oplus B_3^+\;.\nn
 \ea
 The corresponding basis polynomials of ${\mathcal H}_{\Gamma}$ can be written as 
 $|\Gamma,\alpha;l,n\rangle$, where $\alpha$ runs from $1$ to $d_\Gamma$, the dimension of $\Gamma$, and $n$ runs from 
 $1$ to $N(\Gamma,l)$, the total number of occurrences of $\Gamma$ in the decomposition of the space ${\mathcal H}_l$. The relation between
 $|\Gamma,\alpha;l,n\rangle$ and ${\mathcal Y}_{lm}({\mathbf x})$ is written as
 \bd
 |\Gamma,\alpha;l,n\rangle=\sum_{m=-l}^{l}c^l_{\Gamma,\alpha,n;m}{\mathcal Y}_{lm}({\mathbf x})\;,\quad
 {\mathcal Y}_{lm}({\mathbf x})=\sum_{\Gamma}\sum_{n=1}^{N(\Gamma,l)}\sum_{\alpha=1}^{d_\Gamma}c^{l*}_{\Gamma,\alpha,n;m}|\Gamma,\alpha;l,n\rangle\;,
 \ed
 where the coefficients $c^l_{\Gamma,\alpha,n;m}$ form a $(2l+1)\times(2l+1)$ matrix $C_l$ satisfying $C_l^\dagger C_l=1$. 
  We list $|\Gamma,\alpha;l,n\rangle$ in terms of ${\mathcal Y}_{lm}({\mathbf x})$ for $l=0,1$ in Table~{\ref{tab:basis}}.
\begin{table}[htb]
\begin{center}
\begin{tabular}{|c|c|c|}
  \hline
  $|\Gamma,\alpha;l,n\rangle$ & in terms of ${\mathcal Y}_{lm}({\mathbf x})$ & in terms of polynomials \\
  \hline
  $|A^+,1;0,1\rangle$ & ${\mathcal Y}_{00}({\mathbf x})$ & $\frac{1}{\sqrt{4\pi}}$ \\
  \hline
  $|B_1^-,1;1,1\rangle$ & $-\frac{1-i}{2}{\mathcal Y}_{11}({\mathbf x})+\frac{1+i}{2}{\mathcal Y}_{1\bar{1}}({\mathbf x})$ & $\frac{\sqrt{3}}{\sqrt{8\pi}}(x_1+x_2)$ \\
  \hline
  $|B_2^-,1;1,1\rangle$ & $-\frac{1+i}{2}{\mathcal Y}_{11}({\mathbf x})+\frac{1-i}{2}{\mathcal Y}_{1\bar{1}}({\mathbf x})$ & $\frac{\sqrt{3}}{\sqrt{8\pi}}(x_1-x_2)$ \\
  \hline
  $|B_3^-,1;1,1\rangle$ & ${\mathcal Y}_{10}({\mathbf x})$ & $\frac{\sqrt{3}}{\sqrt{4\pi}}x_3$\\
  \hline
\end{tabular}
\caption{$|\Gamma,\alpha;l,n\rangle$ expanded in terms of ${\mathcal Y}_{lm}({\mathbf x})$ for $l=0,1$.  The notation $\bar{m}=-m$ is used.}
\label{tab:basis}
\end{center}
\end{table}

The matrix $D^{(l)}(\hat{R})$ can be diagonalized through
 \bd
 C_{l}D^{(l)}C_{l}^{\dagger}=
 \left(
    \begin{array}{ccc}
     D^{(\Gamma_1)} & & \\
      & D^{(\Gamma_2)} & \\
      & & \ddots \\
    \end{array}
 \right)\;,
 \ed
 where $D^{(\Gamma)}(\hat{R})$ is the matrix representation of $D_{2h}$ for the irrep $\Gamma$. It gives the action of the group element $\hat{R}\in \mathbb G$
 in terms of linear transformations on the vector space ${\mathcal H}_\Gamma$ as follows
 \bd
 \hat{R}|\Gamma,\alpha;l,n\rangle=D^{(\Gamma)}_{\alpha\beta}(\hat{R})|\Gamma,\beta;l,n\rangle\;.
 \ed
 From inspecting Eqs.~(\ref{eq:expansion_Bessel_L}) and (\ref{eq:symmetric}), it can be seen
 that the matrix $\mathcal{M}^{\mathbf d}_{lm,l'm'}$ transforms as
 \bd
 \sum_sD^{(l)}_{ms}(\hat{R})\mathcal{M}^{\mathbf d}_{ls,l'm'}=
 \sum_{s'}\mathcal{M}^{\mathbf d}_{lm,l's'}D^{(l')}_{s'm'}(\hat{R})\;,\;\;\forall\;\hat{R}\in{\mathbb G}\;.
 \ed
 According to Schur's lemma, we then have
 \be
 \label{eq:Schur}
 C^l_{\Gamma,\alpha,n;m}\mathcal{M}^{\mathbf d}_{lm,l'm'}C^{l'*}_{\Gamma',\alpha',n';m'}=
 \delta_{\Gamma,\Gamma'}\delta_{\alpha,\alpha'}{M}^{\mathbf d}_{ln,l'n'}(\Gamma)\;.
 \ee

 Using the projection operator 
 \bd 
 \hat{P}_{\Gamma,\alpha}=\frac{d_\Gamma}{N_G}\sum_{\hat{R}\in{\mathbb G}}D^{(\Gamma)}_{\alpha\alpha}(\hat{R})^*\hat{R}\;,
 \quad N_G=\sum_{\hat{R}\in{\mathbb G}}1\;,
 \ed 
 we can project
 $\phi_{CM}(\mathbf x)$ to the $\Gamma$ representation through $\phi_{CM}^{\Gamma,\alpha}(\mathbf x)=\hat{P}_{\Gamma,\alpha}\phi_{CM}(\mathbf x)$.
 Following from the orthogonality theorem
 \bd
 \frac{d_\Gamma}{N_G}\sum_{\hat{R}\in{\mathbb G}}D^{(\Gamma)}_{\alpha\beta}(\hat{R})^*D^{(\Gamma')}_{\alpha'\beta'}(\hat{R})=\delta_{\Gamma\Gamma'}\delta_{\alpha\alpha'}\delta_{\beta\beta'}\;,
 \ed 
 one can prove that
 \ba
 \label{eq:wave_func_gamma}
 \phi_{CM}^{\Gamma,\alpha}(\mathbf x)&=&\sum_{l,m,m'}\sum_{n=1}^{N(\Gamma,l)}\nu_{lm}c^{l*}_{\Gamma,\alpha,n;m}
 c^{l}_{\Gamma,\alpha,n;m'}G_{lm'}(\mathbf x,p^2)\nn\\
 &=&\sum_{l,n}\tilde{\nu}_{ln}
 \left\{r^{-l}|\Gamma,\alpha;l,n\rangle n_l(pr)+\sum_{l',n'}
 {M}^{\mathbf d}_{ln,l'n'}(\Gamma)r^{-l'}|\Gamma,\alpha;l',n'\rangle j_{l'}(pr)\right\}
 \ea
 where the coefficients $\tilde{\nu}_{ln}$ are defined as
 \bd
 \tilde{\nu}_{ln}=\frac{(-1)^l}{4\pi}p^{l+1}\sum_{m=-l}^{l}\nu_{lm}c^{l*}_{\Gamma,\alpha,n;m}\;.
 \ed
 
 \section{Finite-size formulae}
 In Eq.~(\ref{eq:wave_func_gamma}), we have expanded $\phi_{CM}^{\Gamma,\alpha}(\mathbf x)$ in terms of the singular ${\mathbf d}$-periodic solutions of the Helmholtz equation. In the exterior region, it can also be expanded in terms of the spherical harmonics $Y_{lm}(\hat{\mathbf x})$
 and spherical Bessel functions $j_l(pr)$ and $n_l(pr)$ as
 \be
 \label{eq:wave_func_gamma1}
 \phi_{CM}^{\Gamma,\alpha}(\mathbf x)=\sum_{l,m}b_{lm}Y_{lm}(\hat{\mathbf x})\left(\alpha_l(p)j_l(pr)+\beta_l(p)n_l(pr)\right)\;.
 \ee
 It was proved in Ref.~\cite{Luscher:1990ux} that there exists a unique eigenfunction of the Hamiltonian in the infinite volume 
 which coincides with $\phi_{CM}^{\Gamma,\alpha}(\mathbf x)$ in the exterior region. As a result,
 the coefficients $\alpha_l(p)$ and $\beta_l(p)$ in Eq.~(\ref{eq:wave_func_gamma1}) can be related to $l$-th 
 wave scattering phase $\delta_l(p)$ through
 \bd
 e^{2i\delta_l(p)}=\frac{\alpha_l(p)+i\beta_l(p)}{\alpha_l(p)-i\beta_l(p)}\;.
 \ed
 Eqs.~(\ref{eq:wave_func_gamma1}) and (\ref{eq:wave_func_gamma}) together determine the coefficients $b_{lm}$ that make the two expansions of $\phi_{CM}^{\Gamma,\alpha}(\mathbf x)$ coincide. Writing $b_{lm}$ as
 $b_{lm}=\sum_n\tilde{b}_{ln}c^l_{\Gamma,\alpha,n;m}$,
 Eq.~(\ref{eq:wave_func_gamma1}) can then be written as
 \bd
 \phi_{CM}^{\Gamma,\alpha}(\mathbf x)=\sum_{l,n}\tilde{b}_{ln}r^{-l}|\Gamma,\alpha;l,n\rangle\left(\alpha_l(p)j_l(pr)+\beta_l(p)n_l(pr)\right)\;.
 \ed
 We therefore get the relation between $\tilde{\nu}_{ln}$ and $\tilde{b}_{ln}$ as
 \ba
 &&\tilde{b}_{ln}\beta_l(p)=\tilde{\nu}_{ln}\;,\nn\\
 &&\tilde{b}_{ln}\alpha_l(p)=\sum_{l',n'}\tilde{\nu}_{l'n'}{M}^{\mathbf d}_{l'n',ln}(\Gamma)\;.\nn
 \ea
 This homogenous system has a non-trivial solution of $\tilde{b}_{ln}$ only when the determinant of the coefficient matrix equals zero, which results in
 \be
 \label{eq:FSM_general}
 \mathrm{det}\left[\xi-M^{\mathbf d}(\Gamma)\right]=0\;,\quad \xi_{ln,l'n'}=\delta_{l,l'}\delta_{n,n'}\tan^{-1}\delta_l(p)\;.
 \ee
 If we consider only the lowest angular momentum contribution to Eq.~(\ref{eq:FSM_general}), then the finite-size formulae
 are given by
 \ba
 &&\tan^{-1}\delta_0(p)={M}^{\mathbf d}_{01,01}(A^+)\;,\nn\\
 &&\tan^{-1}\delta_1(p)={M}^{\mathbf d}_{11,11}(\Gamma)\;,\quad\mathrm{for\;\;}\Gamma=B_1^-,\;B_2^-,\;B_3^-\;.\nn
 \ea
 Using Tables~\ref{value_M}, \ref{tab:basis} and Eq.~(\ref{eq:Schur}), we can describe ${M}^{\mathbf d}_{ln,l'n'}(\Gamma)$ in terms of $\mathcal{Z}_{lm}^{\mathbf d}(1;q^2)$
 \ba
&&M^{\mathbf d}_{01,01}(A^+)=(\gamma\pi^{3/2}q)^{-1}\mathcal{Z}^{{\mathbf d}}_{00}\nn\\
&&M^{\mathbf d}_{11,11}(B_1^-)=(\gamma\pi^{3/2}q)^{-1}(\mathcal{Z}^{{\mathbf d}}_{00}-\frac{q^{-2}}{\sqrt{5}}\mathcal{Z}^{{\mathbf d}}_{20}
+i\frac{\sqrt{3}q^{-2}}{\sqrt{10}}(\mathcal{Z}^{{\mathbf d}}_{22}-\mathcal{Z}^{{\mathbf d}}_{2\bar{2}})) \nn\\
&&M^{\mathbf d}_{11,11}(B_2^-)=(\gamma\pi^{3/2}q)^{-1}(\mathcal{Z}^{{\mathbf d}}_{00}-\frac{q^{-2}}{\sqrt{5}}\mathcal{Z}^{{\mathbf d}}_{20}
-i\frac{\sqrt{3}q^{-2}}{\sqrt{10}}(\mathcal{Z}^{{\mathbf d}}_{22}-\mathcal{Z}^{{\mathbf d}}_{2\bar{2}})) \nn\\
&&M^{\mathbf d}_{11,11}(B_3^-)=(\gamma\pi^{3/2}q)^{-1}(\mathcal{Z}^{{\mathbf d}}_{00}+\frac{2q^{-2}}{\sqrt{5}}\mathcal{Z}^{{\mathbf d}}_{20})\;.
 \nn
\ea
 Thus we obtain the finite-size formulae in the MF2 for the irreps $\Gamma=A^+$, $B_1^-$, $B_2^-$ and $B_3^-$.

\section{Conclusion}
The finite-size formulae for a new MF with total momentum $\mathbf P=(2\pi/L)({\mathbf e}_1+{\mathbf e}_2)$ 
are derived in this work. These formulae can be used to calculate the S-wave and P-wave scattering phases.
Using similar procedures to those in this work, one can easily generalize 
the derivation to other MFs and different irreps.

\section{Acknowledgments}
This work is supported by the DFG Sonderforschungsbereich / Transregio SFB/TR-9, 
the DFG project Mu 757/13, the U.S. DOE under Contract No. DE-AC05-06OR23177 and the Grant-in-Aid of the Japanese Ministry of Education (No. 21674002).
X.\ F.\ would like to thank N. Ishizuka and C. Liu for valuable
discussions.

\bibliographystyle{hunsrt}
\bibliography{MF2.bib}

%

\end{document}